\documentclass{aa}  

\usepackage{graphicx}
\usepackage{txfonts}
\usepackage{lipsum}
\usepackage{subcaption}         % necessary for continued figures, example in section 3
\usepackage{lscape}             % to rotate a single page table, example in appendix.
\usepackage{placeins}           % useful with \FloatBarrier, to keep 
\usepackage{gensymb}

\usepackage{natbib}
\bibpunct{(}{)}{;}{a}{}{,} % to follow the A&A style

\newcommand{\hii}{H~{\small II}~}
\usepackage[]{hyperref}
\begin{document}

   \title{The dynamical environment of the high-mass star-forming region G28.288--0.364}

   %\subtitle{Multiplicity and evolutionary status of the associated dust cores}

%%%%%%%%%%%%%%%%%%%%%%%%%%%%%%%%%%%%%%%%
% Please do not include ORCIDs next to author names.
% Only ORCIDs authenticated by individual authors in EDP Sciences editorial system will be taken into account.
% ORCIDs included here will be removed.
%%%%%%%%%%%%%%%%%%%%%%%%%%%%%%%%%%%%%%%%

   \author{Jyotirmoy Dey\inst{1}\fnmsep\thanks{Corresponding Author}
        \and Devendra K. Ojha\inst{1}
        \and Jagadheep D. Pandian\inst{2}
        }

   \institute{Department of Astronomy \& Astrophysics, Tata Institute of Fundamental Research, Mumbai, 400005, India\\
             \email{dey.jyotirmoy04@gmail.com, jyotirmoy.dey@tifr.res.in}
            \and Department of Earth \& Space Sciences, Indian Institute of Space Science and Technology (IIST), Trivandrum 695547, India\\ }

   \date{Received }

% \abstract{}{}{}{}{}
% 5 {} token are mandatory
 
  \abstract
   {Massive stars form within deeply embedded dust cores, and the development of hypercompact and ultracompact \hii regions characterizes their early evolution. Identifying and analyzing such regions is essential for understanding the physical processes that govern massive star formation and the transitions between early evolutionary stages.}
   {We investigate the physical and kinematic properties of the massive star-forming region G28.288--0.364 to constrain the evolutionary stages of embedded \hii regions and the surrounding cores.} 
   {We analyze multiwavelength observations, including uGMRT radio continuum data; archival continuum and radio recombination line data from the GLOSTAR-D survey; high-angular-resolution ALMA Band~3 radio recombination line and 1.36-mm dust continuum data from the ALMAGAL survey; and complementary molecular line tracers. We derive spectral indices, measure linewidths and velocities of the radio recombination line emission, identify compact dust cores using dendrogram analysis, and estimate their physical properties.}
   {The radio continuum emission exhibits a positive spectral index ($\alpha = 0.39 \pm 0.01$), consistent with partially optically thick free-free emission. High-resolution observations resolve the ionized gas into two distinct components with physical sizes of $\sim$ 0.06~pc. Their RRL linewidths ($\sim$ 37 and 32~km~s$^{-1}$, respectively) indicate that one component is in a transitional stage between hypercompact and ultracompact \hii regions, while the other is more evolved. %A large-scale velocity \textbf{shift} is detected in the ionized gas, likely tracing extended bulk motions of the gas.
   The 1.36-mm dust continuum data reveal five dust cores with surface densities consistent with the theoretical threshold for massive star formation.}
   {G28.288--0.364 hosts multiple compact \hii regions and dense cores at different evolutionary stages, including a rare transitional object between the hypercompact and ultracompact phases. These results highlight the complex and sequential nature of massive star formation within clustered environments and demonstrate the importance of high angular resolution multiwavelength observations for resolving the early evolution of massive stars.}

   \keywords{Stars: Massive -- ISM: \hii regions -- ISM: molecules -- Molecular data -- Instrumentation: Interferometers
               }

   \maketitle
   \nolinenumbers

%%%%%%%%%%%%%%%%%%%%%%%%%%%%%%%%%%%%%%%%%%%%%%%%%%%%%%%%%%%%%%
\section{Introduction}\label{sect:intro}
%\lipsum[1]

The formation of massive stars ($M$ $\geq$ 8~M$_{\odot}$) remains one of the least understood domains in modern astrophysics. Although recent works \citep{2007ARA&A..45..481Z,2011MNRAS.416..972D,2014prpl.conf..149T,2023NatAs...7..557B,2023MNRAS.520.2306T,2024ApJ...966...54P,2025ARA&A..63....1B,2025ApJ...986...15T} have shed some light on this topic, the exact physical processes through which massive stars accumulate their mass are still debated. Currently, two main scenarios are proposed to explain this phenomenon, ``competitive accretion'' \citep{1998MNRAS.298...93B,2004MNRAS.349..735B,2007prpl.conf..149B} and ``core accretion'' \citep{2002Natur.416...59M,2003ApJ...585..850M}.

In the ``competitive accretion'' scenario,  a molecular cloud fragments into a group of small cores, each with a mass roughly equal to the thermal Jeans mass. These cores compete to accrete mass from a common reservoir following its gravitational potential well, and the most massive stars are formed towards the center of the well. In contrast, in the ``core accretion'' model, accretion occurs locally. In this scenario, the mass accreted onto the massive star originates locally, at birth, within the core.

Regardless of the uncertainty in their formation scenario, massive stars emit copious amounts of ionizing photons that ionize their surroundings, creating \hii regions. These ionized media expand and evolve with time, transforming from the hypercompact (HC) to ultracompact (UC) to compact \hii regions. Although numerous studies have investigated the properties of UC \hii regions and compact \hii regions \citep{1989ApJS...69..831W,2006ApJS..165..338Q,2011MNRAS.416..972D,2014MNRAS.438.1335R}, there are relatively fewer works that have focused on HC \hii regions \citep{2019A&A...624A.100C,2019MNRAS.482.2681Y,2021A&A...645A.110Y,2023MNRAS.524.4384P,2024MNRAS.533.2005P,2025MNRAS.538.2267P}. This scarcity is largely due to their extremely compact nature (sizes $\lesssim$ 0.03~pc; \citealt{2005IAUS..227..111K}), the high optical depth at frequencies of 1--8~GHz, where most radio surveys have been carried out so far, and their short lifetimes ($\leq$ $10^5$~years; \citealt{2003ApJ...599.1196K}), all of which make their detection and detailed characterization particularly challenging.

In this paper, we present the results of our multi-wavelength study of G28.288--0.364 (RA 18h\,44m\,15.1s, Dec --04$\degree$\,17$'$\,55$\farcs$9; J2000), a high-mass star-forming region that is listed as a candidate UC \hii region in the Co-Ordinated Radio `N' Infrared Survey for High-mass star formation (CORNISH) catalog \citep{2012PASP..124..939H,2013ApJS..205....1P}  with a total flux density of 552.77$\pm$51.90~mJy. \cite{2012ApJ...756...60S} have reported a systemic velocity of 48.7~km~s$^{-1}$ for G28.288--0.364. They have also put this star-forming region at a near kinematic distance of 3.3~kpc, which is also consistent with the distance estimations of \cite{2006ApJ...653.1325S} and \cite{2016ApJ...823...77R}.

\begin{figure}
\centering
\includegraphics[width=0.95\linewidth]{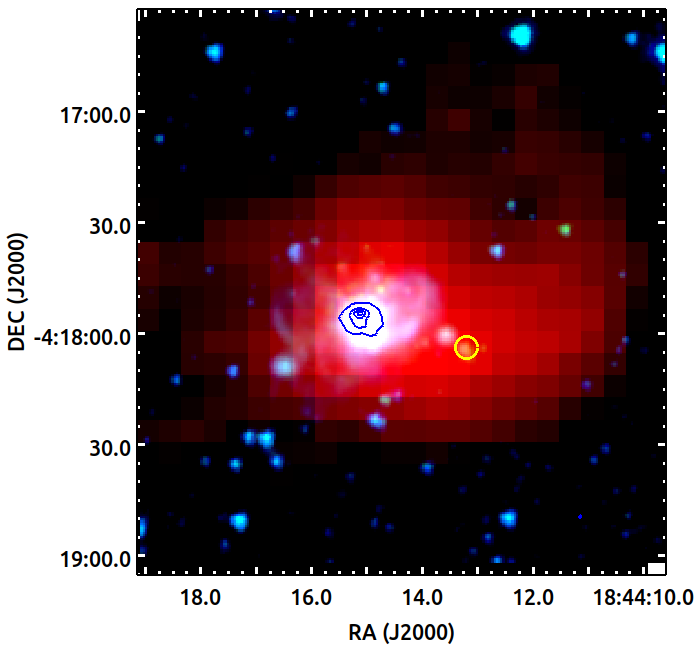}
\caption{Three-colour image showing emission at 870~$\mu$m (red), 4.5~$\mu$m (green), and 3.6~$\mu$m (blue) from the ATLASGAL (870~$\mu$m) and GLIMPSE (4.5~$\mu$m and 3.6~$\mu$m) surveys. The blue contours correspond to the radio continuum emission from the CORNISH survey, starting at the $3\sigma$ level with $\sigma = 0.5$ mJy~beam$^{-1}$. The position of the extended green object (EGO) identified by \cite{2011ApJ...743...56C} is marked by the yellow circle. The emission is displayed on a logarithmic scale to highlight the faint components.
}\label{fig:rgb}
\end{figure}

In addition, \cite{2011ApJ...743...56C} have detected an extended green object (EGO) $\lesssim$ 0.5~pc away (RA: 18h\,44m\,13.2s, Dec: --04$\degree$\,18$'$\,04$\farcs$0; J2000) from G28.288--0.364 (see Fig.~\ref{fig:rgb}). The systemic velocity of this EGO is 49.5~km~s$^{-1}$ \citep{2009ApJ...702.1615C}, which is in good agreement with the systemic velocity of G28.288--0.364. Thus, it is very likely that the EGO and \hii region belong to the same star-forming complex. Cyanopolyyne lines (HC$_3$N, HC$_5$N, and HC$_7$N) have also been detected toward G28.288--0.364 \citep{2016ApJ...830..106T,2018ApJ...866...32T}. The presence of these species indicates that warm carbon-chain chemistry is active in this region. %This, in turn, suggests that G28.288--0.364 may host a ``hot-core'', with an excitation temperature of approximately 100--200 K \citep{2016ApJ...830..106T}.

This manuscript is organized as follows. In Sect.~\ref{sect:data}, we describe the data used in this study from our observations and various archives. The results of our analysis are presented in Sect.~\ref{sect:res}. In Sect.~\ref{sect:dis}, we discuss the broader implications and the global picture associated with G28.288--0.364. Finally, our main conclusions are summarized in Sect.~\ref{sect:conclude}.

%%%%%%%%%%%%%%%%%%%%%%%%%%%%%%%%%%%%%%%%%%%%%%%%%%%%%%%%%%%%%%

\section{Observations and archival data}\label{sect:data}

\subsection{uGMRT observations}
The radio observations of G28.288--0.364 were carried out using the upgraded Giant Metrewave Radio Telescope (uGMRT) with the GMRT Wideband Backend (GWB) correlator (proposal code: 40\_100). The correlator was configured with a total bandwidth of 200 MHz, centered at 1350 MHz (Band-5), and divided into 8192 spectral channels. At this frequency, the uGMRT provides a native angular resolution of approximately $2''$ and a largest recoverable angular scale of about $7'$ in Band-5.

The data were reduced and analyzed using the NRAO Common Astronomy Software Applications (CASA; \citealt{2007ASPC..376..127M}) package. Initial flagging of corrupted data was performed using the \texttt{flagdata} task. Gain solutions were then derived and applied to the calibrator sources, followed by a second round of data inspection and flagging. Final bandpass and gain solutions were subsequently computed and applied to the target source. A detailed description of the calibration and imaging procedures is presented in \cite{2024A&A...689A.254D}. The resulting continuum image achieved a final $1\sigma$ rms noise level of 103~$\mu$Jy~beam$^{-1}$, with a synthesized beam size of $3\farcs02 \times 2\farcs27$.

\subsection{Archival data}
\subsubsection{GLOSTAR survey}
The uGMRT observations are complemented by data from the A global view on star formation (GLOSTAR; \citealt{2021A&A...651A..85B}) survey. The GLOSTAR survey was conducted with the Karl G. Jansky Very Large Array (VLA) in its D and B configurations. The correlator setup covers full-polarization continuum emission over the 4--8~GHz frequency range, using two 1 GHz-wide sub-bands centered at 4.7 and 6.9~GHz. In addition to the continuum, the survey includes observations of the 6.7~GHz methanol maser line, the 4.8~GHz formaldehyde line, and seven radio recombination lines (RRLs). Zero-spacing information is provided by complementary observations with the 100-m Effelsberg radio telescope.

We have used radio continuum images and recombination line data cubes from the VLA D-configuration of the GLOSTAR survey (hereafter GLOSTAR-D; \citealt{2019A&A...627A.175M}). The GLOSTAR-D continuum products include a single image at an effective frequency of 5.79 GHz, as well as eight sub-band images that enable spectral index measurements. Because the angular resolution depends on both observing frequency and position on the sky, all continuum images were convolved to a common circular restoring beam of $18''$. Owing to their intrinsically weak emission, all detected RRLs were stacked to improve the signal-to-noise ratio of the final spectral cubes \citep{2024A&A...689A..81K}. The resulting stacked RRL cube has a velocity resolution of 5~km~s$^{-1}$ and an angular resolution of $25''$.

\subsubsection{CORNISH survey}
Radio continuum data from the Co-Ordinated Radio `N' Infrared Survey for High-mass star formation (CORNISH; \citealt{2012PASP..124..939H}) have also been utilized for our analysis. CORNISH is a sensitive, high-angular-resolution 5-GHz radio continuum survey designed to investigate massive star-forming regions, while also providing a valuable resource for a broad range of astrophysical studies. The survey was carried out with the VLA in its B configuration, producing maps with an angular resolution of approximately $1\farcs5$. With a typical rms noise level of $\lesssim 0.4$~mJy~beam$^{-1}$, the data are well suited for identifying UC \hii regions throughout the Galaxy, in addition to detecting other classes of radio-emitting sources.

\subsubsection{ALMAGAL survey}
Dust continuum and molecular line emission data at the millimeter/sub-millimeter wavelengths have been retrieved from the Atacama Large Millimeter/submillimeter Array (ALMA) FITS Archive\footnote{Part of the data are retrieved from the JVO portal (\url{http://jvo.nao.ac.jp/portal}) operated by the NAOJ.} (Project ID: 2019.1.00195.L; PI: Sergio Molinari). These observations were conducted as a part of the ALMA Evolutionary Study of High Mass Protocluster Formation in the Galaxy (ALMAGAL; \citealt{2025A&A...696A.149M}) survey. The ALMAGAL survey is a large ALMA program approved in Cycle 7. The survey targets over 1000 compact, dense molecular clumps distributed across the Milky Way and spanning a wide range of evolutionary stages. The full sample is widely distributed throughout the Galactic plane. Observations were carried out with ALMA in Band~6, covering frequencies between 217 and 221~GHz (corresponding to a wavelength of $\sim$1.3~mm). Detailed descriptions of the observing strategy, data calibration, and imaging procedures are provided in the dedicated ALMAGAL data reduction paper \citep{2025A&A...696A.150S}.

\subsubsection{Ancillary ALMA data}
In addition to the ALMAGAL data, we made use of ALMA Band~3 observations of the H63$\delta$ RRL (Project ID: 2023.1.00467.S; PI: Kotomi Taniguchi) to investigate the kinematics of the ionized gas. The observations were carried out within a frequency range of 95.805--110.386~GHz, providing a velocity resolution of 0.214~km~s$^{-1}$ and an angular resolution of $0\farcs997$. The data reach a typical rms sensitivity of 0.484~mJy~beam$^{-1}$ per channel, enabling detailed analysis of the velocity structure and line widths of the ionized gas.

\subsubsection{ATLASGAL survey}
We made use of archival 870~$\mu$m continuum data from the APEX Telescope Large Area Survey of the Galaxy (ATLASGAL; \citealt{2009A&A...504..415S}) to investigate the large-scale distribution of cold dust emission associated with the target region. The survey was carried out using the Atacama Pathfinder Experiment (APEX) telescope and provides an angular resolution of $\sim19\farcs2$ at 870~$\mu$m. ATLASGAL data are particularly well suited for tracing dense clumps and filamentary structures in the Galactic plane, enabling the identification of the parental molecular environment and the distribution of dense gas associated with star-forming regions.

\subsubsection{GLIMPSE survey}
We also made use of archival infrared data from the Galactic Legacy Infrared Mid-Plane Survey Extraordinaire (GLIMPSE; \citealt{2003PASP..115..953B}) obtained with the Infrared Array Camera (IRAC) onboard the Spitzer Space Telescope. The survey provides imaging of the Galactic plane in the 3.6, 4.5, 5.8, and 8.0~$\mu$m bands with high angular resolution and sensitivity. The GLIMPSE data were used to examine the infrared morphology of G28.288--0.364 and to trace infrared dark clouds (IRDCs) associated with this region.

%The $^{12}$CO, $^{13}$CO, and C$^{18}$O $J$ = 1--0 transitions data from the FOREST Unbiased Galactic plane Imaging survey with the Nobeyama 45-m telescope (FUGIN; \citealt{2017PASJ...69...78U}) have been utilized to study the molecular environment surrounding G28.288--0.364. The FUGIN survey is a large legacy project conducted with the Nobeyama Radio Observatory’s 45-m telescope equipped with the multi-beam FOREST receiver. FUGIN maps extensive portions of the Galactic plane in the $^{12}$CO, $^{13}$CO, and C$^{18}$O $J$ = 1--0 transitions simultaneously. This survey attains an angular resolution of $\sim 20''$, significantly finer than previous large-scale CO surveys, enabling detailed studies of molecular gas distribution, kinematics, and physical properties from large-scale structures down to filaments, clumps, and cores.

%%%%%%%%%%%%%%%%%%%%%%%%%%%%%%%%%%%%%%%%%%%%%%%%%%%%%%%%%%%%%%

\section{Results}\label{sect:res}

\subsection{Radio continuum emission}\label{sect:radio_cont}

In addition to our uGMRT observations, G28.288--0.364 is covered by several Galactic plane surveys. Table~\ref{tab:radio_flux} summarizes the integrated radio continuum flux densities ($S_\nu$) measured at the $3\sigma$-level from the corresponding continuum images. Using these measurements, we derived a spectral index ($\alpha$; $S_\nu \propto \nu^{\alpha}$) of $0.39 \pm 0.01$, indicative of partially optically thick free–free emission. This is in agreement with the in-band spectral index ($\alpha_{\rm in-band} = 0.16 \pm 0.02$) reported in the GLOSTAR-D radio source catalog \citep{2019A&A...627A.175M} that also indicates the ionized gas to be partially optically thick.

\begin{table}[h!]
\caption{\label{tab:radio_flux} The uGMRT and archival radio continuum flux densities.}
\centering
\begin{tabular}{lcc}
\hline\hline
Survey/Telescope Name & $\nu$ & $S_\nu$ \\
 & (GHz) & (Jy) \\
\hline
uGMRT & 1.35 & 0.397 $\pm$ 0.005 \\
THOR$^{\rm a}$ & 1.44 & 0.389 $\pm$ 0.019 \\
CORNISH$^{\rm b}$ & 4.86 &  0.543 $\pm$ 0.014 \\
GLOSTAR-D$^{\rm c}$ & 5.79 & 0.699 $\pm$ 0.003 \\
GLOSTAR-D$+$Effelsberg$^{\rm d}$ & 5.85 & 0.720 $\pm$ 0.018 \\
\hline
\end{tabular}
\tablefoot{The $S_\nu$ values are estimated after applying a 3$\sigma$-level threshold to the corresponding radio continuum maps. a $=$ \cite{2016A&A...595A..32B,2016A&A...588A..97B}, b $=$ \cite{2013ApJS..205....1P}, c $=$ \cite{2019A&A...627A.175M,2024A&A...689A.196M}, d $=$ \cite{2026A&A...706A.230G}}
\end{table}

Interestingly, the high-resolution ($\sim 1\farcs5$) 5~GHz radio continuum map from the CORNISH survey revealed two distinct continuum emission components (hereafter CORNISH-A and CORNISH-B) that are not detected in other surveys, including our uGMRT observations. This discrepancy is likely attributable to the poorer angular resolution of those surveys, which may dilute or blend the compact emission features.

\begin{figure}[t!]
\centering
\includegraphics[width=0.49\textwidth]{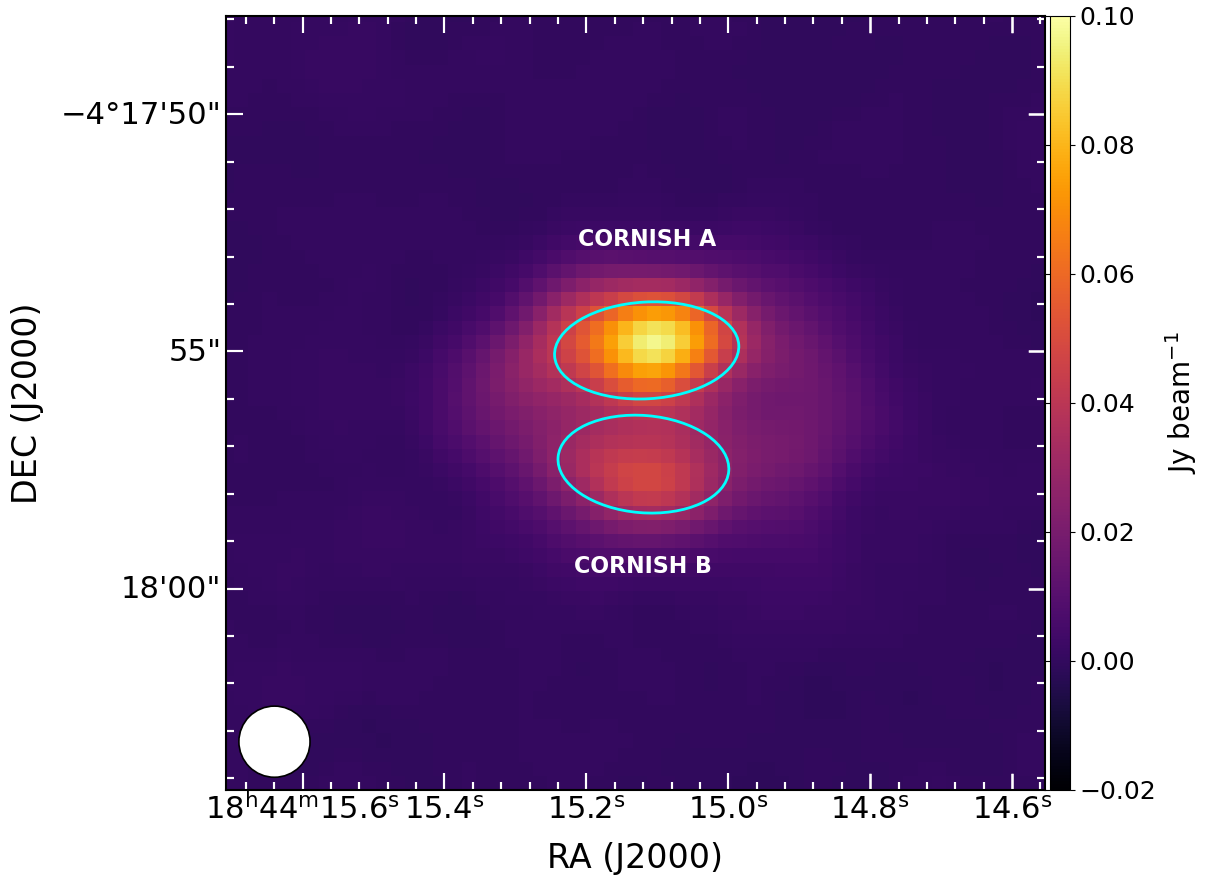}
\caption{Image showing the radio continuum map from the CORNISH survey. The cyan ellipses indicate the fitted components, CORNISH-A and CORNISH-B. The synthesized beam ($\sim 1\farcs5$) is shown in the bottom-left corner of the figure.
}\label{fig:r_cont}
\end{figure}

We employed the \texttt{imfit} task within the CASA package to estimate the angular extents of these components (see Fig.~\ref{fig:r_cont}). The deconvolved source sizes returned by \texttt{imfit} are $3\farcs89 \pm 0\farcs28 \times 2\farcs04 \pm 0\farcs16$ for CORNISH-A and $3\farcs61 \pm 0\farcs20 \times 2\farcs05 \pm 0\farcs12$ for CORNISH-B. Adopting the distance to G28.288--0.364 given in Sect.~\ref{sect:intro}, these angular dimensions correspond to physical sizes of $0.062 \pm 0.004$~pc $\times$ $0.033 \pm 0.002$~pc for CORNISH-A and $0.058 \pm 0.003$~pc $\times$ $0.033 \pm 0.002$~pc for CORNISH-B.

According to the size-based classification scheme of \cite{2005IAUS..227..111K}, both CORNISH-A and CORNISH-B fall in the regime of \hii regions evolving from the HC to the UC phase. This interpretation is further supported by the detection of a positive spectral index over the 1--5~GHz frequency range, consistent with partially optically thick free–free emission. However, a robust confirmation of this evolutionary stage requires an analysis of the ionized gas kinematics. In particular, measurements of the RRL full width at half maximum (FWHM) can distinguish between HC and UC \hii regions, as HC regions typically exhibit broader lines ($\gtrsim$ 40~km~s$^{-1}$) due to strong turbulence, pressure broadening, or expansion \citep{2004ApJ...605..285S,2008ApJ...672..423K}. In contrast, UC regions show comparatively narrower line widths.

\begin{figure}[t!]
\centering
\includegraphics[width=0.99\linewidth]{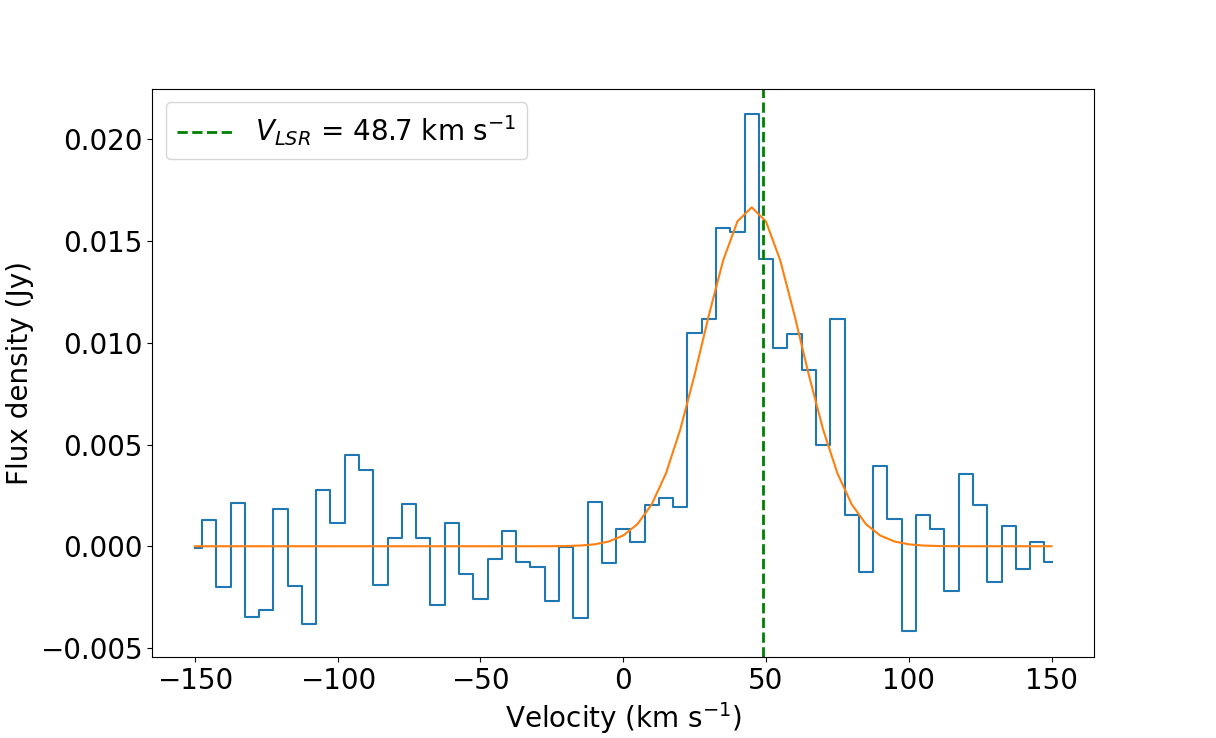}
\includegraphics[width=0.99\linewidth]{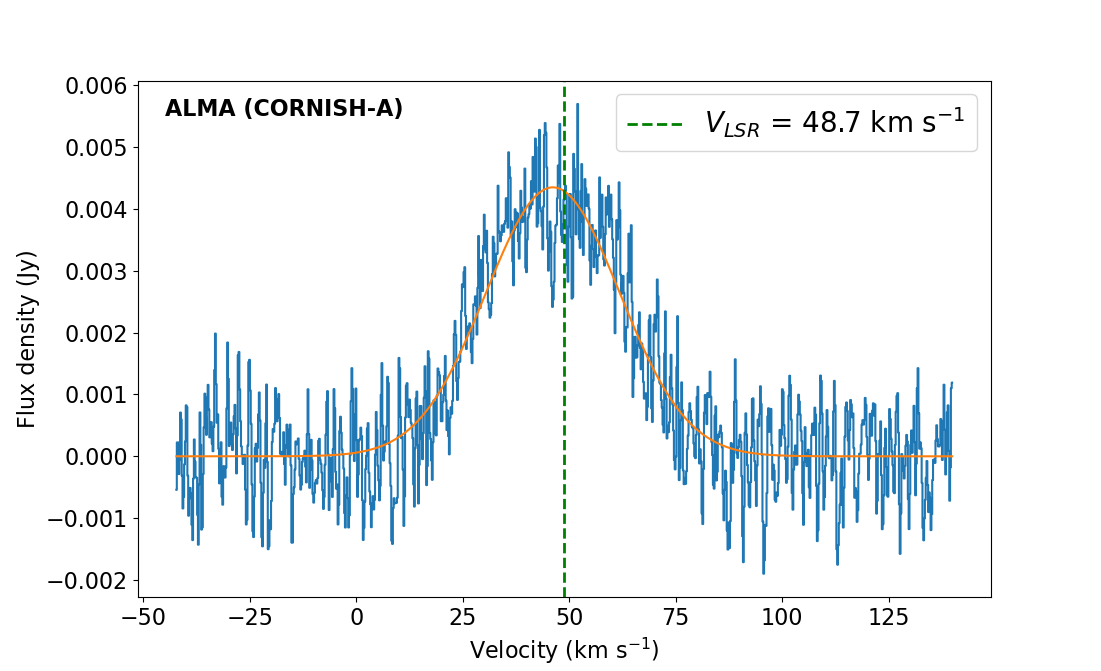}
\includegraphics[width=0.99\linewidth]{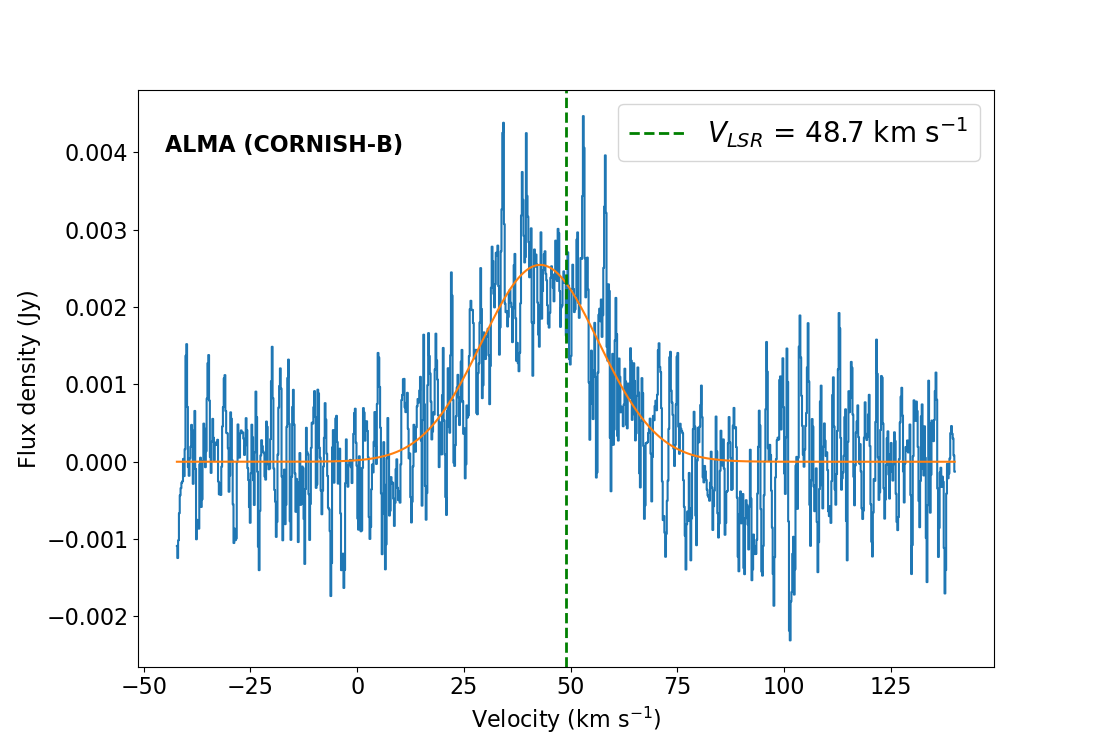}
\caption{Average GLOSTAR-D RRL spectrum toward G28.288$-$0.364 (top), extracted from a circular region with a diameter equal to one synthesized beam of the GLOSTAR-D RRL map. Similar RRL spectra are also extracted toward CORNISH-A (middle) and CORNISH-B (bottom) using the ALMA Band 3 H63$\delta$ RRL data. The orange curves show the Gaussian fits to the line profiles, while the green dotted lines mark the systemic velocity reported by \cite{2012ApJ...756...60S}.
}\label{fig:glos_rrl}
\end{figure}

\subsection{Radio recombination line emission}\label{sect:radio_line}

The RRLs are a powerful diagnostic tool for the physical and kinematic properties of \hii regions. Arising from electron–ion recombination in ionized gas, RRLs directly trace the conditions within photoionized nebulae and enable measurements of systemic velocities, electron temperatures, FWHMs, and ionized gas kinematics \citep{2015MNRAS.450.2025A,2018A&A...616A.107K}. When combined with radio continuum observations, RRL data allow robust constraints on the nature of the ionizing sources and the evolutionary state of \hii regions, as well as the identification of thermal and non-thermal emission components \citep{2006ApJS..165..338Q,2018ApJS..234...33A}. In this section, we analyze the detected RRL emission to investigate the physical properties and kinematics of the ionized gas associated with the target region.

Both the GLOSTAR-D and ALMA Band~3 data showed radio recombination line emission toward G28.288--0.364. However, in contrast to the high-angular-resolution CORNISH radio continuum map, the stacked GLOSTAR-D RRL data showed only a single RRL-emitting component. This is most likely a consequence of the relatively coarse angular resolution of the GLOSTAR-D RRL maps ($\sim25''$), which blends the compact ionized structures detected in higher-resolution observations.

Consequently, we extracted an average RRL spectrum (see the top panel of Fig.~\ref{fig:glos_rrl}) from a circular region with a diameter equal to one synthesized beam of the GLOSTAR-D RRL map, chosen to fully encompass the $3\sigma$-level radio continuum emission seen in the CORNISH map. The fitted GLOSTAR-D RRL spectrum peaks at a central velocity of  $44.99 \pm 1.44$~km~s$^{-1}$ and has a FWHM of $40.33 \pm 3.39$~km~s$^{-1}$. Accounting for the $1\sigma$ uncertainty, the measured FWHM is consistent with the lower end of the FWHM distribution observed toward the Galactic HC \hii regions \citep{2004ApJ...605..285S,2019MNRAS.482.2681Y,2021A&A...645A.110Y}.

% \begin{figure}
% \centering
% \includegraphics[width=0.95\linewidth]{rrl_m1_glostar.png}
% \caption{Intensity-weighted velocity (moment~1) map derived from the GLOSTAR-D RRL data. The contour levels are the same as those shown in Fig.~\ref{fig:rgb}. The synthesized beam ($\sim 25''$) is indicated in the bottom-left corner of the figure.
% }\label{fig:moment1}
% \end{figure}

Next, we extracted average spectra from the ALMA Band~3 H63$\delta$ RRL data. These observations provide a substantially higher angular resolution ($\sim 1''$) than the GLOSTAR-D RRL data, enabling us to determine the RRL properties of CORNISH-A and CORNISH-B individually (see the middle and bottom panels of Fig.~\ref{fig:glos_rrl}). For CORNISH-A, the fitted spectrum yields a central velocity of $46.16 \pm 0.29$~km~s$^{-1}$ and an FWHM of $37.07 \pm 0.69$~km~s$^{-1}$, whereas for CORNISH-B, the central velocity and FWHM are $42.86 \pm 0.48$~km~s$^{-1}$ and $32.33 \pm 1.13$~km~s$^{-1}$, respectively.

Given the estimated sizes of CORNISH-A and CORNISH-B (Sect.~\ref{sect:radio_cont}), the fitted RRL FWHMs suggest that CORNISH-A is at an intermediate evolutionary stage between a HC and a UC \hii region, whereas CORNISH-B appears to be at a more evolved stage.

%In addition, the moment~1 (intensity-weighted velocity) map generated from the GLOSTAR-D RRL cube reveals a large velocity \textbf{shift} of $\gtrsim 10$~km~s$^{-1}$ across the direction nearly perpendicular to the line connecting CORNISH-A and CORNISH-B (see Fig.~\ref{fig:moment1}). The physical origin of such a large velocity \textbf{shift} in the ionized gas remains unclear. Similar \textbf{shifts} reported in previous studies have been attributed to a variety of mechanisms, including expansion or infall motions \citep{2018A&A...611A..99K}, stellar-driven outflows \citep{2013A&A...556A.107K}, cloud–cloud collisions \citep{2022ApJ...925...60D}, and rotation \citep{2021ApJ...921..176B}. 

%We note, however, that this velocity \textbf{shift} is not apparent in the higher angular resolution ALMA Band~3 data. This may indicate that the observed \textbf{shift} arises from more extended ionized emission that is resolved out in the interferometric observations, or reflects large-scale bulk motions rather than compact rotational or localized dynamical structures.

\subsection{Dust continuum emission}\label{sect:dust}

We used the ALMAGAL 1.36~mm dust continuum maps to identify and characterize compact cores associated with G28.288--0.364. Core extraction was performed using the \texttt{astrodendro}\footnote{\texttt{astrodendro} is a \texttt{Python} package for computing dendrograms of astronomical data (\url{http://www.dendrograms.org/})} package together with the CASA \texttt{imfit} task. The dendrogram algorithm \citep{2008ApJ...679.1338R} decomposes the emission into a hierarchical set of structures and provides a quantitative description of the intensity distribution in star-forming regions. In this framework, the smallest structures that are not further subdivided (leaves) are considered candidate star-forming cores.

The identification of leaves is controlled by three parameters: (i) a minimum intensity threshold ($min\_value$) of 3$\sigma$, where $\sigma$ is the rms noise of the map; (ii) a minimum intensity contrast ($min\_delta$) of 1$\sigma$, which sets the required significance for a structure to be treated as independent; and (iii) a minimum number of pixels ($min\_npix$), chosen to be equal to the synthesized beam area expressed in pixel units, ensuring that only spatially resolved structures are selected. For each identified core, the algorithm returns the centroid position, deconvolved major and minor axis sizes, and position angle.

Although the dendrogram technique is well-suited for investigating the hierarchical organization of emission within dense clumps, it cannot be used directly to derive fragment masses. This limitation arises because the dendrogram formalism quantifies only the relative intensity contrast of a leaf with respect to its parent node, that is, the height of a peak above the corresponding saddle point, rather than the total integrated flux associated with the structure.

To overcome this limitation, we employed the CASA task \texttt{imfit}, using the parameters derived from the dendrogram analysis as initial guesses for the fitting. All fit parameters were allowed to vary. The fitting is performed within a bounding box whose dimensions were determined by the aperture sizes of the corresponding leaves identified in the dendrogram analysis. The derived source parameters, including the peak position (RA, DEC), deconvolved major and minor axes (FWHM$_{\rm maj}$, FWHM$_{\rm min}$), position angle (PA), peak intensity ($F_{\rm peak}$), and integrated flux density ($F_{\rm int}$), are listed in Table~\ref{tab:dust_cont}.

\begin{figure*}
\centering
\includegraphics[width=0.51\linewidth]{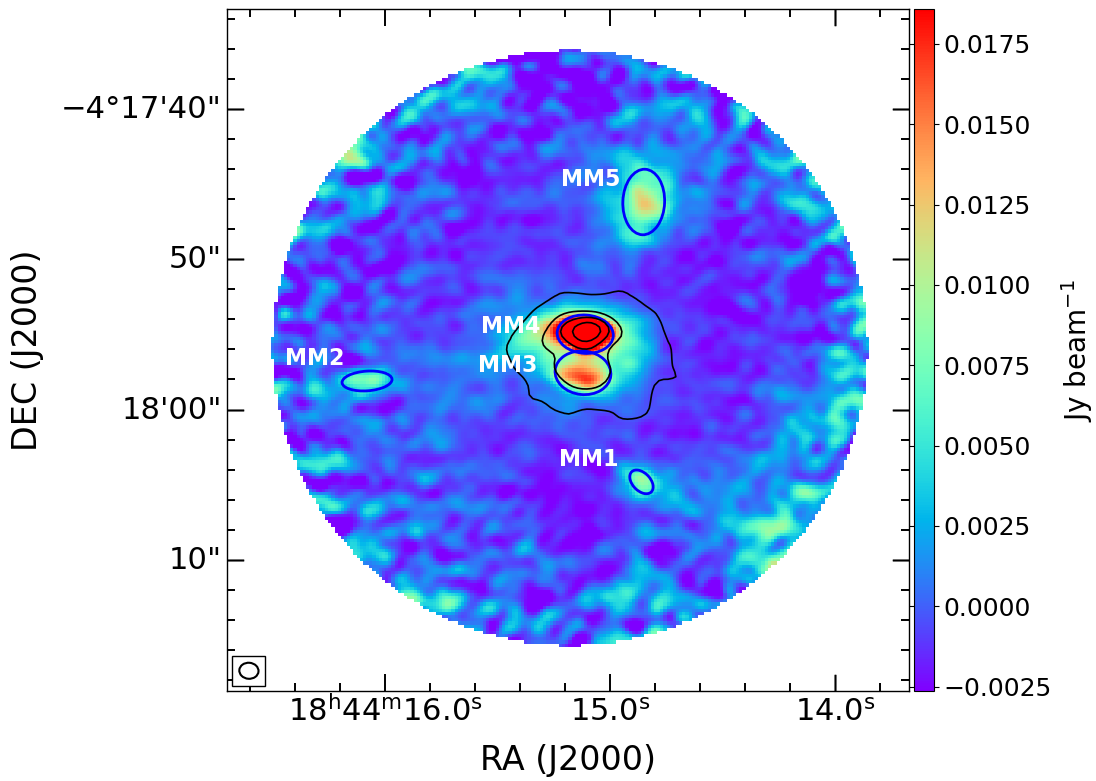}
\includegraphics[width=0.47\linewidth]{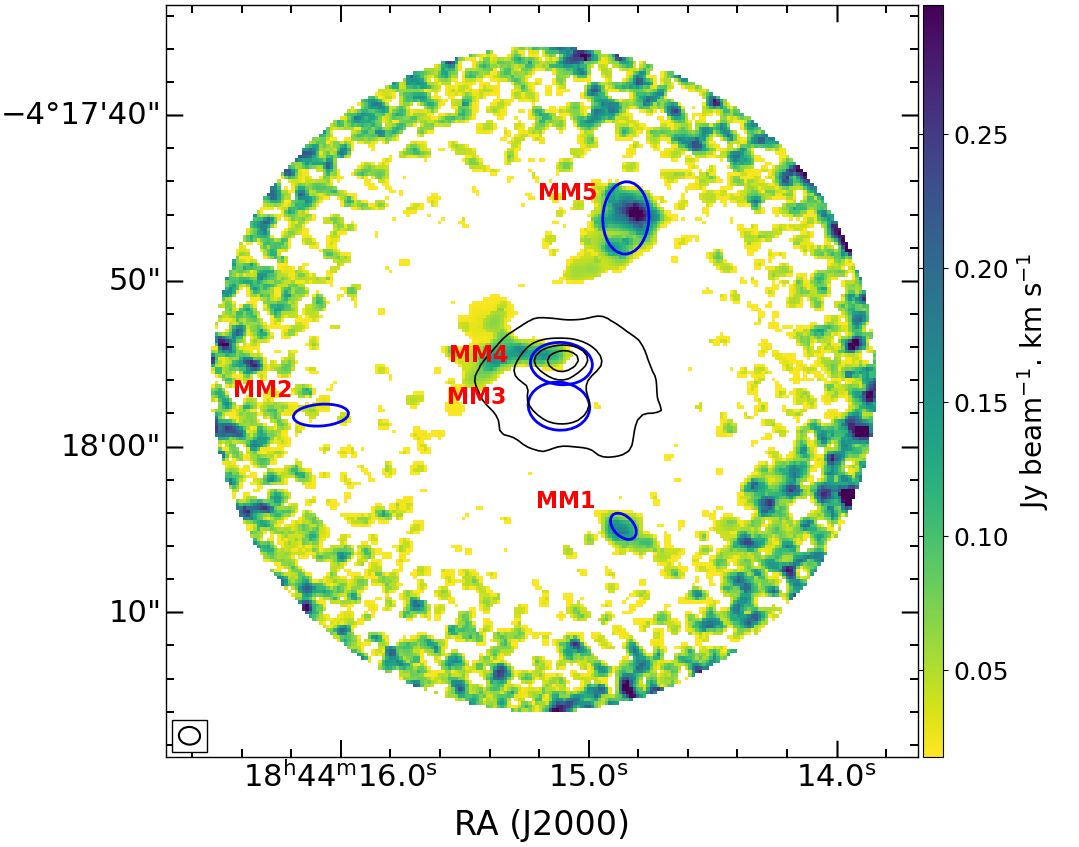}
\caption{Dust cores (blue ellipses) identified in Sect.~\ref{sect:dust} overlaid on the 1.36~mm dust continuum map (left) and the DCN (3--2) moment~0 map (right) from the ALMAGAL survey. The contour levels are the same as those shown in Fig.~\ref{fig:rgb}. The synthesized beams are indicated in the bottom-left corner of the respective panels.}\label{fig:dcn}
\end{figure*}

To avoid including spurious detections, we retained only sources satisfying $F_{\rm peak} > 5\sigma$. In addition, we rejected cores with poorly fitted morphologies based on visual inspection of the 1.36-mm dust continuum map overlaid with the identified leaf structures. Such cases typically correspond to filamentary features or diffuse emission, characterized by an aspect ratio greater than three between the major and minor axes. Table~\ref{tab:dust_cont} lists only those cores that satisfy the above criteria (also see the left panel of Fig.~\ref{fig:dcn}).

\begin{table*}[h!]
\caption{\label{tab:dust_cont} Physical parameters of the detected cores.}
\centering
\begin{tabular}{ccccccccccc}
\hline\hline
Core & RA & DEC & FWHM$_{\rm maj}$ & FWHM$_{\rm min}$ & PA & $F_{\rm peak}$ & $F_{\rm int}$ & $M_{\rm core}$ & $\Sigma_{\rm core}$ & n \\
 & (J2000) & (J2000) & (arcsec) & (arcsec) & (deg) & (mJy~beam$^{-1}$) & (mJy) & (M$_\odot$) & (g~cm$^{-2}$) & ($\times10^{6}$ \\
 &  &  &  &  &  &  &  &  &  & $\text{cm}^{-3}$) \\
\hline
MM1 & 18:44:14.86 & --04:18:04.81 & 1.87 & 1.19 & 44.0 & 9.1 & 24.7 & 4.06 & 1.89 & 5.50 \\
MM2 & 18:44:16.08 & --04:17:58.10 & 3.33 & 1.31 & 94.0 & 7.7 & 34.4 & 5.65 & 1.35 & 2.79 \\
MM3 & 18:44:15.12 & --04:17:57.56 & 3.72 & 2.89 & 84.0 & 15.8 & 141.0 & 23.15 & 2.24 & 2.95 \\
MM4 & 18:44:15.11 & --04:17:54.99 & 3.74 & 2.55 & 86.0 & 36.7 & 295.0 & 48.43 & 5.28 & 7.40 \\
MM5 & 18:44:14.85 & --04:17:46.20 & 4.36 & 2.78 & 177.7 & 10.8 & 109.1 & 17.91 & 1.54 & 1.91 \\
\hline
\end{tabular}
%\tablefoot{}
\end{table*}

Next, assuming that the dust emission is optically thin, we derived the isothermal core masses using the following expression

\begin{equation}
    M_{\rm core} = \frac{\delta_{\rm gd}\,F_{\rm int}\,D^2}{\kappa_\nu\,B_\nu(T_{\rm d})}
\end{equation}
where $M_{\rm core}$ denotes the core mass, $\delta_{\rm gd}$ is the gas-to-dust mass ratio, $F_{\rm int}$ is the integrated flux density, and $D$ is the distance to the source. $\kappa_\nu = \kappa_{\nu_0}\,\left(\frac{\nu}{\nu_0}\right)^{\beta}$ is the dust opacity, where $\beta$ is the dust emissivity index, adopted as 2.0 \citep{2013A&A...554A..42R} in our calculations. $B_\nu(T_{\rm d})$ is the Planck function evaluated at the dust temperature $T_{\rm d}$.

For our calculations, we considered $\delta_{\rm gd} = 100$, $\kappa_{\nu_0}$ = 0.899~cm$^2$~g$^{-1}$ at 1300~$\mu$m or 1.3~mm for a MRN-distribution of grain sizes (for the diffuse interstellar medium of gas density $=10^6$~cm$^{-3}$) with thin ice mantles \citep{1994A&A...291..943O}, and $T_{\rm d}$ = 30.9~K for the entire ATLASGAL clump \citep{2022MNRAS.510.3389U}.

The mass surface densities ($\Sigma_{\rm core}$) and number densities ($n$) were also derived using the following expressions

\begin{equation}
    \Sigma_{\rm core} = \frac{M_{\rm core}}{\pi\,R_{\rm core}^2}
\end{equation}

\begin{equation}
    n \approx \frac{\Sigma_{\rm core}}{2\,R_{\rm core}\,\mu\,m_{\rm H}}
\end{equation}
where the mean molecular weight ($\mu$)~$\approx 2.8$, $m_{\rm H} = 1.67\times10^{-24}$~g, $R_{\rm core}$ is the effective core radius, defined as half of the geometric mean of FWHM$_{\rm maj}$ and FWHM$_{\rm min}$ evaluated at the distance of G28.288--0.364. Our estimated masses, mass surface densities, and number densities are reported in Table~\ref{tab:dust_cont}.

Based on the surface density threshold for massive star formation ($\gtrsim 1$ g cm$^{-2}$; \citealt{2003ApJ...585..850M}), all cores listed in Table~\ref{tab:dust_cont} satisfy the criterion for forming massive stars. However, the high-angular-resolution ALMAGAL observations indicate that MM1 and MM2 do not contain sufficient mass to form massive stars yet.

However, it's important to mention that the reported 1.36~mm (220.6~GHz) continuum flux density in Table~\ref{tab:dust_cont} is expected to arise from a combination of thermal dust emission and free-free emission from ionized gas. Since the turnover frequency is not constrained by either the available radio continuum data or dedicated modeling of the ionized component, a reliable decomposition of these contributions is not possible. We therefore estimate the free-free contribution by adopting the highest observed radio frequency as the turnover frequency and extrapolating the measured flux density to 220.6~GHz, assuming optically thin free-free emission with a spectral index of $\alpha=-0.1$. This approach provides a conservative estimate of the minimum free-free contribution to the 1.36~mm continuum flux density.

For the dust cores in G28.288$-$0.364, the radio continuum flux densities were derived from the CORNISH survey, whose angular resolution is comparable to that of the ALMAGAL 1.36~mm continuum images. No CORNISH radio continuum counterparts are detected toward MM1, MM2, and MM5. Thus, it's assumed that the cold dust emission is not contaminated by the free-free emission, and no correction is required for these cores. Whereas MM3 and MM4 are associated with the \hii regions CORNISH-B and CORNISH-A, with measured flux densities of $157.8\pm5.5$~mJy and $253.6\pm13.0$~mJy, respectively.

Extrapolating these values to 220.6~GHz under the assumption of optically thin free-free emission, we estimate that ionized gas accounts for approximately 76\% and 59\% of the 1.36~mm flux densities reported in Table~\ref{tab:dust_cont} for MM3 and MM4 (corresponding to approximately 108~mJy and 173~mJy, respectively). Consequently, attributing the full 1.36~mm flux densities in Table~\ref{tab:dust_cont} entirely to thermal dust emission would overestimate the masses of MM3 and MM4. Subtracting the extrapolated free-free contribution from the reported 1.36~mm flux densities, the resulting dust masses of MM3 and MM4 are reduced by factors of 4.23 and 2.43, respectively, yielding revised masses of 5.47 and 19.93~M$_\odot$.

This behavior is again consistent with the evolutionary stages of MM3: the detection of a more evolved \hii region (CORNISH-B) indicates that the central object has already progressed to a stage where it actively ionizes and begins to disperse its surrounding envelope, reducing the reservoir of cold dust. In comparison, MM4 exhibits a comparatively smaller correction to its mass, suggesting a less advanced stage of \hii region (CORNISH-A) or a denser surrounding envelope.

\subsection{Molecular line emission}\label{sect:molecule}

\begin{table}[t!]
\caption{\label{tab:line_param} Properties of the analyzed spectral lines.}
\centering
\begin{tabular}{p{1cm}p{1cm}p{1cm}p{1cm}p{1cm}p{1cm}}
\hline\hline
\multicolumn{1}{c}{Mole-} & Quantum & \multicolumn{1}{c}{Freq-} & Core & $T_{\rm rot}$ & \multicolumn{1}{c}{N} \\
\multicolumn{1}{c}{cule} & numbers & \multicolumn{1}{c}{uency} &  &  &  \\
 &  & (GHz) &  & (K) & (cm$^{-2}$) \\
\hline
DCN & 3--2 & 217.238 & MM1 & -- & 7.34$\times10^{12}$  \\
 &  &  & MM2 & -- & 4.26$\times10^{12}$ \\
 &  &  & MM3 & -- & 1.00$\times10^{12}$ (LL) \\
 &  &  & MM4 & -- & 2.32$\times10^{12}$ \\
 &  &  & MM5 & -- & 9.73$\times10^{12}$ \\
 \hline
H$_2$CO & 3$_{(0,3)}$--2$_{(0,2)}$ & 218.222 & MM1 & 48 & 2.31$\times10^{13}$ \\
 & 3$_{(2,2)}$--2$_{(2,1)}$ & 218.476 & MM2 & 84 & 1.02$\times10^{13}$  \\
 & 3$_{(2,1)}$--2$_{(2,0)}$ & 218.760 & MM3 & 199 & 1.00$\times10^{13}$ (LL) \\
 &  &  & MM4 & 80 & 1.00$\times10^{13}$ (LL) \\
 &  &  & MM5 & 200 & 3.39$\times10^{14}$ \\
\hline
HC$_3$N & 24--23 & 218.325 & MM1 & -- & 1.98$\times10^{13}$ \\
 &  &  & MM2 & -- & 3.62$\times10^{13}$ \\
 &  &  & MM3 & -- & 1.54$\times10^{12}$ \\
 &  &  & MM4 & -- & 7.18$\times10^{13}$ \\
 &  &  & MM5 & -- & 9.77$\times10^{12}$ \\
\hline
C$^{18}$O & 2--1 & 219.560 & MM1 & -- & 5.06$\times10^{16}$ \\
 &  &  & MM2 & -- & 1.14$\times10^{16}$ \\
 &  &  & MM3 & -- & 2.14$\times10^{15}$ \\
 &  &  & MM4 & -- & 2.06$\times10^{15}$ \\
 &  &  & MM5 & -- & 2.19$\times10^{16}$ \\
\hline
${}^{13}$CO & 2--1 & 220.399 & MM1 & -- & 9.72$\times10^{16}$ \\
 &  &  & MM2 & -- & 4.02$\times10^{15}$ \\
 &  &  & MM3 & -- & 1.39$\times10^{16}$ \\
 &  &  & MM4 & -- & 1.51$\times10^{16}$ \\
 &  &  & MM5 & -- & 7.37$\times10^{15}$ \\
\hline
SO & 6$_5$--5$_4$ & 219.949 & MM1 & -- & 1.60$\times10^{14}$ \\
 &  &  & MM2 & -- & 1.16$\times10^{13}$ \\
 &  &  & MM3 & -- & 2.36$\times10^{12}$ \\
 &  &  & MM4 & -- & 5.28$\times10^{12}$ \\
 &  &  & MM5 & -- & 2.63$\times10^{14}$ \\
\hline
\end{tabular}
\tablefoot{LL = lower limit of the input parameter}
\end{table}

\begin{figure}
\centering
\includegraphics[width=\linewidth]{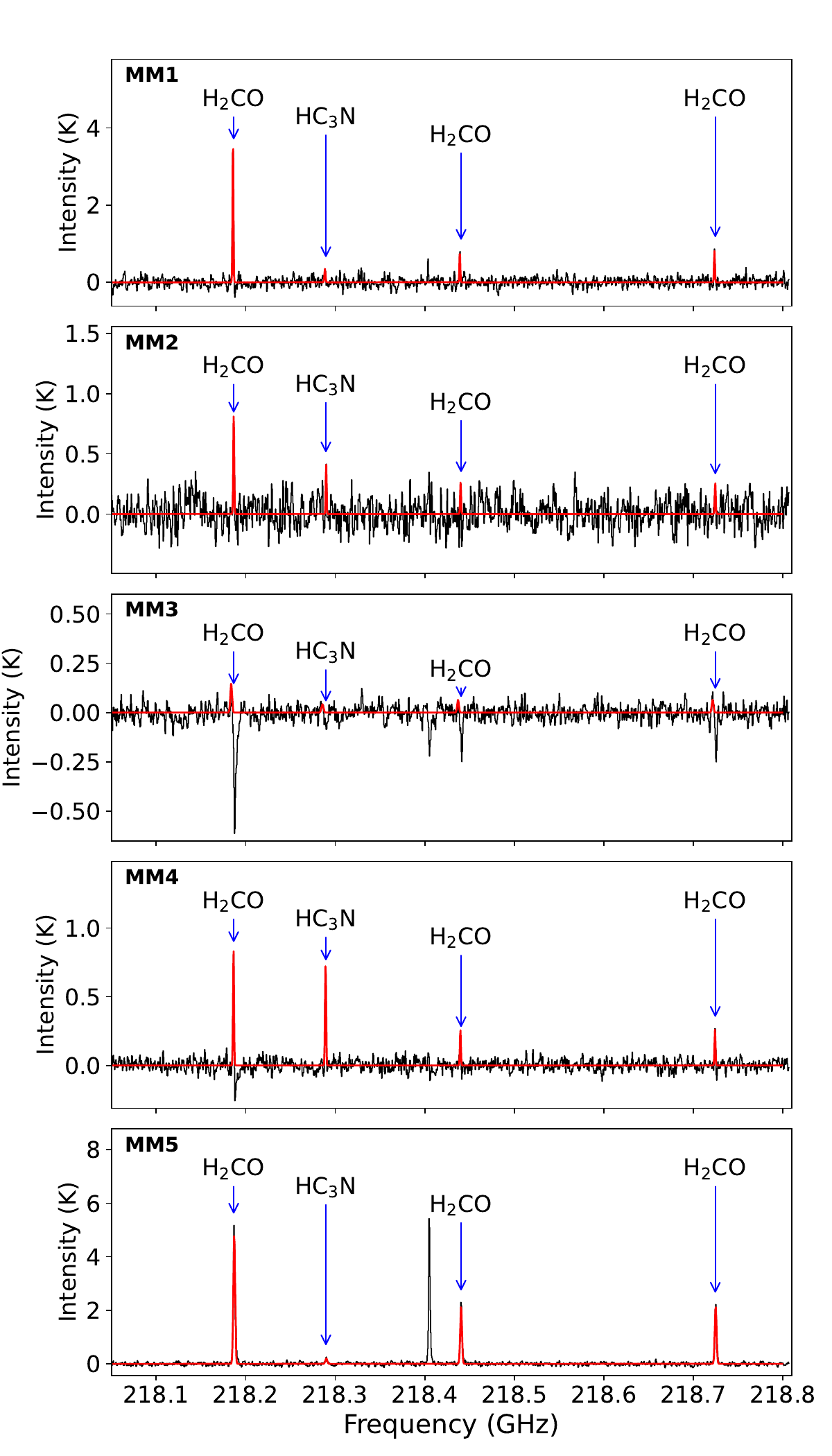}\caption{ H$_2$CO transitions within the ALMAGAL spectral setup.}\label{fig:tran}
\end{figure}

Several molecular transitions were detected toward G28.288--0.364 using the spectral line data from the ALMAGAL survey. Line identification was performed by comparing the observed rest frequencies with entries from the Cologne Database for Molecular Spectroscopy \citep{2001A&A...370L..49M,2005JMoSt.742..215M} and the Jet Propulsion Laboratory Molecular Spectroscopy Catalog \citep{1998JQSRT..60..883P} using the spectral analysis package Centre d'Analyse Scientifique de Spectres Instrumentaux et Synthétiques (CASSIS; \citealt{2015sf2a.conf..313V}).

The molecular emission was modeled using the eXtended CASA Line Analysis Software Suite (XCLASS; \citealt{2017A&A...598A...7M}), which solves the one-dimensional radiative transfer equation under the assumption of local thermodynamic equilibrium (LTE). Given the high number densities associated with the detected cores (see Table~\ref{tab:dust_cont}), with $n \gtrsim 10^6$~cm$^{-3}$ in the vicinity of the dust cores, the LTE approximation is expected to be valid for the observed transitions \citep{2015PASP..127..266M}. XCLASS simultaneously fits the observed spectra using Gaussian line profiles while accounting for optical depth effects, line blending, and contributions from multiple molecular species.

Within XCLASS, each molecular species can be represented by one or more emission and/or absorption components. The free parameters describing each component include the source size ($\theta_{\rm source}$), rotational temperature ($T_{\rm rot}$), column density ($N$), linewidth ($\Delta v$), and velocity offset relative to the systemic velocity ($v_{\rm off}$). The best-fit parameters were obtained through $\chi^2$ minimization using the optimization routines implemented in XCLASS. Table~\ref{tab:line_param} summarizes the detected molecular transitions and their corresponding modeled parameters.

A reliable determination of the rotational temperature, $T_{\rm rot}$, requires the detection of multiple transitions from the same molecular species. In this work, we used formaldehyde (H$_2$CO) as a molecular thermometer to probe the temperature structure of the region, since H$_2$CO exhibits several strong and predominantly optically thin transitions within the ALMAGAL spectral setup (see Fig.~\ref{fig:tran}). Formaldehyde is known to be an effective tracer of relatively low-temperature gas with $T_{\rm kin} \lesssim 100$~K \citep{1993ApJS...89..123M}. However, under conditions of high density and temperature, the H$_2$CO $(3_{0,3}-2_{0,2})$ transition can become optically thick, making temperature estimates based on H$_2$CO line ratios unreliable \citep{2012A&A...545A..51R,2019A&A...631A.142G}. In such cases, methyl cyanide (CH$_3$CN) transitions are generally preferred as temperature tracers. 

Unfortunately, no CH$_3$CN transitions are detected toward G28.288--0.364. This is consistent with the results of \citet{2017ApJ...844...68T,2019ApJ...881...57T}, who reported a lower abundance of complex organic molecules in G28.288--0.364 compared to other sources in their sample. Nevertheless, under the assumption of LTE, the kinetic temperature of the gas can be approximated by the rotational temperature, that is, $T_{\rm kin} \approx T_{\rm rot}$, and our $T_{\rm rot}$ estimates yielded high temperatures ($\gtrsim 150$~K), and possibly saturate for MM3 and MM5, whereas the derived temperatures for MM1, MM2, and MM4 appear to be well constrained and reliable. %The high temperatures inferred for MM3 and MM5, therefore, suggest that these cores may harbor ``hot-core'' activity, in agreement with the predictions of \citet{2016ApJ...830..106T}.
In addition, the H$_2$CO spectra toward MM3 and MM4 show absorption-like features. Further analysis confirms that these arise from large-scale negative emission at the source position, a common artefact in interferometric observations.

We have also overplotted all detected cores on the moment~0 (velocity-integrated intensity) map constructed from the DCN~$(3-2)$ line emission obtained as part of the ALMAGAL survey (see the right panel of Fig.~\ref{fig:dcn}). The DCN~$(3-2)$ has been proposed as a tracer of both early protostellar evolution and dense gas; however, its usefulness as a ``hot-core'' diagnostic remains debated \citep{2015ApJ...804...37L,2024ApJ...963...12T}. Nonetheless, owing to its typically optically thin nature, DCN provides a valuable probe of cold, dense gas in star-forming regions and protoplanetary disks \citep{2023A&A...678A.194C,2023A&A...669A.137H}. Significant DCN~$(3-2)$ emission was detected toward MM1 and MM5, suggesting that they are likely to be early-stage protostellar cores. In contrast, MM3 and MM4 exhibited little to no DCN~$(3-2)$ emission, likely reflecting the destruction of DCN molecules due to heating and enhanced radiation fields associated with the formation of \hii regions. %This, in turn, suggests that out of MM3 and MM5, the latter one is more likely to host a ``hot-core'' as discussed in the previous paragraph.

Surprisingly, MM2 showed no significant DCN~$(3$--$2)$ emission despite the absence of any radio continuum counterpart. This non-detection may be attributed to a combination of factors, such as elevated gas temperatures suppressing deuterium fractionation, chemical youth or insufficient time for DCN enrichment, depletion of DCN onto dust grains in dense regions, or unfavorable excitation conditions leading to weak line emission. A detailed chemical and radiative transfer analysis would be required to disentangle these effects; however, such an investigation is beyond the scope of the present paper.

It is also worth noting that CORNISH-A and CORNISH-B, as well as their associated host cores MM4 and MM3, exhibit comparable sizes within the respective fitting uncertainties.

\section{Discussions}\label{sect:dis}

\subsection{Outflow activity}\label{sect:outflow}

\cite{2018ApJS..235....3Y} conducted a systematic and unbiased survey of molecular outflows toward 325 clumps selected from the ATLASGAL survey, which is an 870~$\mu$m continuum survey that uniformly maps the inner Galactic plane ($\lvert l \rvert \leq 60\degree$, $\lvert b \rvert \leq 1\rlap{.}\degree5$). To identify and characterize the outflow lobes, they made use of $^{13}$CO and C$^{18}$O ($J = 3-2$) data from the $^{13}$CO/C$^{18}$O ($J = 3-2$) Heterodyne Inner Milky Way Plane Survey (CHIMPS; \citealt{2016ApJ...823...77R}) that covers $28\degree \lesssim l \lesssim 46\degree$ and $\lvert b \rvert \leq 0\rlap{.}\degree5$ in the inner Galaxy and was carried out with the James Clerk Maxwell Telescope.

They first extracted the $^{13}$CO and C$^{18}$O ($J = 3-2$) spectra at the position of the peak 870~$\mu$m continuum emission of the ATLASGAL clump associated with G28.288--0.364. The identification of blue- and redshifted outflow lobes was then carried out using the method described by \cite{2014MNRAS.444..566D}, which is based on the techniques originally introduced by \cite{2007A&A...464.1015V} and \cite{2004A&A...417..615C}. Applying this procedure to the extracted molecular line spectra, \cite{2018ApJS..235....3Y} derived the velocity intervals corresponding to the blue- and redshifted outflow lobes of the parent ATLASGAL clump and estimated a mass entrainment rate of 26.60$\times$10$^{-4}$~M$_\odot$~yr$^{-1}$. Such a high mass entrainment rate is indicative of ongoing massive star formation activity within the parent ATLASGAL clump.

\begin{figure}
\centering
\includegraphics[width=0.95\linewidth]{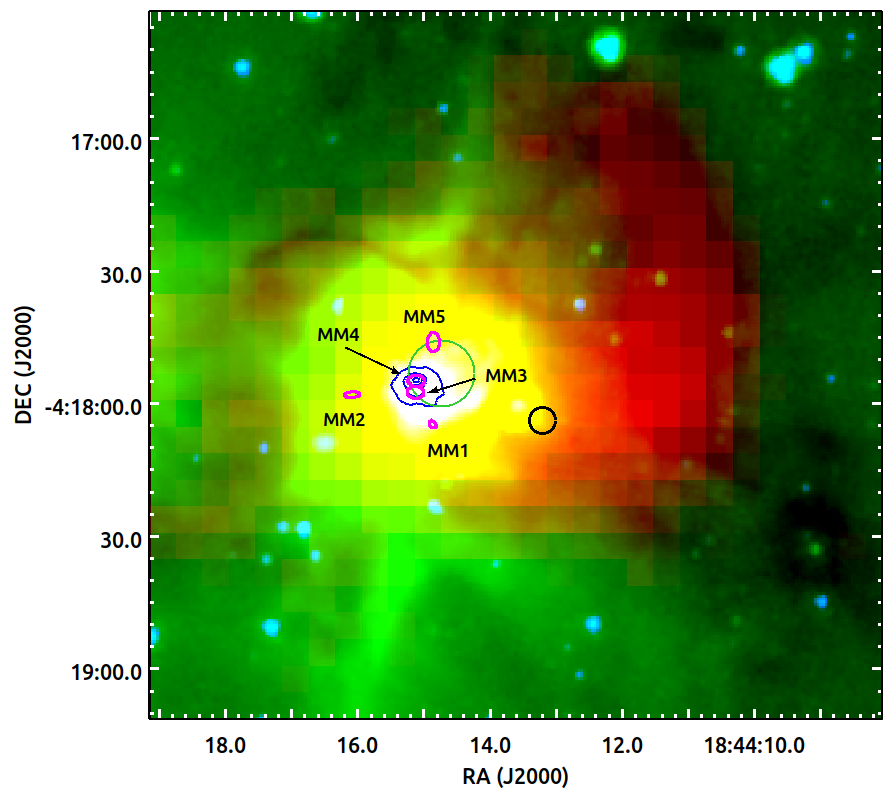}
\caption{Three-colour image showing emission at 870~$\mu$m (red), 8.0~$\mu$m (green), and 3.6~$\mu$m (blue) from the ATLASGAL (870~$\mu$m) and GLIMPSE (8.0~$\mu$m and 3.6~$\mu$m) surveys. The dark feature in the 8.0~$\mu$m image corresponds to the IRDC reported by \cite{2006ApJ...653.1325S} and \cite{2012ApJ...756...60S}. The same region also shows emission at 870~$\mu$m in the ATLASGAL map, tracing the parent ATLASGAL clump associated with G28.288$-$0.364. The blue contours are the same as those shown in Fig.~\ref{fig:rgb}. The magenta ellipses indicate the dust cores identified in Sect.~\ref{sect:dust}. The green circle shows one beam ($\sim15''$) of the CHIMPS survey centered on the peak position of the ATLASGAL clump, while the black circle marks the EGO previously reported by \cite{2011ApJ...743...56C}.
}\label{fig:atlasgal_clump}
\end{figure}

However, as described in Sect.~\ref{sect:dust}, five dust cores are identified in the 1.36-mm continuum map from the ALMAGAL survey. Of these, MM3, MM4 (the host cores of CORNISH-B and CORNISH-A, respectively), and MM5 lie within a single CHIMPS beam ($\sim 15''$) centered on the peak position of the 870-$\mu$m continuum emission of the associated ATLASGAL clump (see Fig.~\ref{fig:atlasgal_clump}).

This spatial coincidence suggests that the outflow lobes identified by \cite{2018ApJS..235....3Y} may arise from MM3, MM4, or MM5. The peak velocities of the $^{13}$CO and C$^{18}$O ($J = 3-2$) lines (48.3 and 47.8~km~s$^{-1}$, respectively) are consistent with the systemic velocity of G28.288--0.364 (48.7~km~s$^{-1}$; \citealt{2012ApJ...756...60S}), and therefore with MM3 and MM4, as well as with MM5 (48.2~km~s$^{-1}$, derived from a Gaussian fit to the DCN ($3-2$) line). Given the angular resolution of the CHIMPS data, however, the outflow emission cannot be unambiguously attributed to any individual core.

Moreover, the $^{13}$CO and C$^{18}$O ($J = 1-0$) data from the FOREST Unbiased Galactic plane Imaging survey with the Nobeyama 45-m telescope (FUGIN; \citealt{2017PASJ...69...78U}) have an angular resolution of $\sim 20''$, comparable to that of the CHIMPS survey. As a result, the FUGIN data are subject to similar beam dilution and blending effects, and thus do not provide additional constraints for isolating the driving source of the outflow.

Although the high-resolution ALMAGAL data products include $^{13}$CO and C$^{18}$O ($J = 2-1$) line emission, these interferometric data cubes, particularly the C$^{18}$O ($J = 2-1$) transition, are affected by spatial filtering of extended emission. Consequently, they are not well-suited for reliably tracing the large-scale outflow structures in this region. Therefore, within the limitations of the available datasets, the identification of the specific dust core responsible for the outflow reported by \cite{2018ApJS..235....3Y} remains uncertain.

\subsection{On the evolutionary nature of CORNISH-A and CORNISH-B}\label{sect:evolution}

Based on our analysis in Sect.~\ref{sect:radio_cont} and \ref{sect:radio_line}, and adopting the classification criteria of \cite{2005IAUS..227..111K}, \cite{2004ApJ...605..285S}, and \cite{2008ApJ...672..423K}, which define HC and UC \hii regions as having characteristic sizes $\lesssim$ 0.03~pc and $\lesssim$ 0.1~pc, respectively, and RRL linewidths of $\gtrsim$ 40~km~s$^{-1}$ and $\approx$ 30~km~s$^{-1}$, CORNISH-A appears to be in a transitional or intermediate stage between the HC and UC \hii phases, whereas, CORNISH-B appears to be in a more evolved UC phase.

Such intermediate objects are intrinsically rare. Only recently, the Search for Clandestine Optically Thick Compact \hii Regions (SCOTCH; \citealt{2023MNRAS.524.4384P,2024MNRAS.533.2005P,2025MNRAS.538.2267P}) has identified a total of 48 new HC and intermediate \hii regions. Before these studies, only 28 such objects had been reported in the literature.

However, distance measurements play an important role in estimating the physical size and, in turn, affect the determination of the evolutionary stage of a \hii region. As mentioned in Sect.~\ref{sect:intro}, previous studies, in particular, \cite{2006ApJ...653.1325S}, have placed G28.288--0.364 at a near distance of 3.3~kpc by morphologically matching the $^{13}$CO ($J = 1-0$) emission from the Galactic Ring Survey with the mid-infrared extinction features observed in the Midcourse Space Experiment (MSX; \citealt{1998ApJ...494L.199E,2006ApJ...639..227S}) survey toward IRDCs in this Galactic longitude range. Subsequent work by \cite{2012ApJ...756...60S}, which employed the high-density tracer N$_2$H$^+$ ($JF_1F$ = 123 -- 012) to determine systemic velocities, further supports this distance estimate.

Furthermore, we compared the extinction features observed in the 8-$\mu$m map from the GLIMPSE survey with the 870-$\mu$m emission in the ATLASGAL map and found a close spatial correspondence between the parent ATLASGAL clump of G28.288--0.364 and the IRDC identified by \cite{2006ApJ...653.1325S} and \cite{2012ApJ...756...60S} at this Galactic longitude (see Fig.~\ref{fig:atlasgal_clump}).

The consistency between the kinematic analyses and the observed morphological agreement between mid-infrared extinction and submillimeter dust emission therefore strengthens the adoption of the near distance of 3.3~kpc for G28.288--0.364. Nevertheless, we note that any systematic uncertainty in the distance would directly scale the derived physical sizes and could influence the inferred evolutionary classification of the associated \hii regions.

%%%%%%%%%%%%%%%%%%%%%%%%%%%%%%%%%%%%%%%%%%%%%%%%%%%%%%%%%%%%%%

\section{Conclusions}\label{sect:conclude}

We have presented a multiwavelength analysis of the massive star-forming region G28.288--0.364 using radio continuum and RRL data from uGMRT, GLOSTAR-D, and ALMA Band~3, together with 1.36-mm dust continuum data from ALMAGAL. Our main results can be summarized as follows:

\begin{enumerate}

\item The 1--6~GHz radio continuum integrated flux densities yield a positive spectral index of $\alpha = 0.39 \pm 0.01$, consistent with partially optically thick free-free emission. High-resolution 5~GHz data resolve the ionized emission into two compact components, CORNISH-A and CORNISH-B, with physical sizes of $\sim$0.06~pc.

\item Analysis of the H63$\delta$ RRL reveals FWHM values of $37.1 \pm 0.7$~km~s$^{-1}$ for CORNISH-A and $32.3 \pm 1.1$~km~s$^{-1}$ for CORNISH-B. Combined with their physical sizes, these linewidths suggest that CORNISH-A is in a transitional stage between HC and UC \hii regions, whereas CORNISH-B is more consistent with a comparatively evolved UC \hii region.

%\item The GLOSTAR-D moment~1 map shows a large-scale velocity \textbf{shift} ($\gtrsim 10$~km~s$^{-1}$) in the ionized gas that is not recovered in the higher-resolution ALMA data. This indicates that the \textbf{shift} likely traces extended ionized emission or large-scale bulk motions rather than compact rotational structures.

\item The 1.36-mm dust continuum data resolve five cores (MM1--MM5) with masses ranging from $\sim$4 to 48~M$_\odot$ and mass surface densities exceeding 1~g~cm$^{-2}$. MM3 and MM4 spatially coincide with CORNISH-B and CORNISH-A, respectively, and exhibit comparable sizes within uncertainties.

\item We also note that, for MM3 and MM4, the embedded \hii regions contribute significantly to their 1.36~mm dust continuum flux densities through free-free emission from the ionized gas. Hence, after correcting for this contamination, the corresponding masses are reduced by factors of 4.23 and 2.43, respectively, and the updated mass range of the five dust cores changes to $\sim$4 to 20~M$_\odot$.

\item Several molecular transitions were detected toward the region and modeled under the assumption of LTE using XCLASS. The derived rotational temperatures indicate significant temperature variations among the detected cores.

\item Temperature estimates based on H$_2$CO suggest elevated temperatures ($\gtrsim 150$~K) toward MM3 and MM5, whereas the temperatures derived for MM1, MM2, and MM4 remain comparatively well constrained. %The high inferred temperatures toward MM3 and MM5 are indicative of possible ``hot-core'' activity.

\item DCN~(3--2) emission is detected toward MM1 and MM5, suggesting early-stage protostellar activity, while MM3 and MM4 show little or no DCN emission, likely due to chemical processing associated with embedded \hii\ regions. MM2 lacks both DCN emission and radio continuum, indicating a potentially distinct evolutionary or chemical state. Further chemical and radiative transfer analysis will be required to clarify its nature.

%\item Combining the temperature estimates and DCN emission properties, MM5 appears to be the strongest candidate for harboring a ``hot-core'' within G28.288--0.364.

\item Previous single-dish studies report a high mass entrainment rate toward the parent ATLASGAL clump, indicative of active massive star formation. However, due to limited angular resolution, the driving source of the outflow cannot be uniquely associated with any individual ALMAGAL core.

\end{enumerate}

Adopting the near kinematic distance of 3.3~kpc, supported by both molecular line velocities and the spatial correspondence between mid-infrared extinction and submillimeter dust emission, we conclude that G28.288--0.364 hosts multiple compact and chemically diverse cores at different evolutionary stages. In particular, the coexistence of a transitional (HC-to-UC) and a more evolved UC \hii\ region within the same clump highlights the complex and sequential nature of massive star formation in clustered environments.

Future high-sensitivity and combined interferometric plus single-dish RRL and molecular line observations will be essential to disentangle the large-scale ionized gas kinematics and to identify the driving source of the molecular outflow in this region.

\begin{acknowledgements}
      We thank the anonymous referee for their comments, which have helped to improve this manuscript. J.D. and D.K.O. acknowledge the support of the Department of Atomic Energy, Government of India, under Project Identification No. RTI 4012. J.D. also thanks Dr. Anindya Saha from the Kavli Institute for Astronomy and Astrophysics (KIAA) at Peking University for the fruitful discussions on the molecular line analysis. We thank the staff of the uGMRT that made these observations possible. uGMRT is run by the National Centre for Radio Astrophysics of the Tata Institute of Fundamental Research. This work (partially) uses information from the GLOSTAR database at \url{http://glostar.mpifr-bonn.mpg.de} supported by the MPIfR, Bonn. Additionally, the SIMBAD database, operated at CDS, Strasbourg, France, has been utilized. This research has also made use of resources such as NASA's Astrophysics Data System and CDS's VizieR catalog access tool.
\end{acknowledgements}
% Please note that we have included the references below in
% order to compile the document, but we ask you to:

\bibliographystyle{aa} % style aa.bst
\bibliography{ref.bib} % your references Yourfile.bib

%%%%%%%%%%%%%%%%%%%%%%%%%%%%%%%%%%%%%%%%%%%%%%%%%%%%%%%%%%%%%%%
% Appendices must be placed after   \end{thebibliography}
% They will be placed automatically on a new page.
%%%%%%%%%%%%%%%%%%%%%%%%%%%%%%%%%%%%%%%%%%%%%%%%%%%%%%%%%%%%%%%
\begin{appendix}

\end{appendix}

\end{document}